\documentstyle[aps,preprint]{revtex}

\draft
\tighten
\begin{document}
\title{Field-Induced Transition in the S=1 Antiferromagnetic Chain with Single-Ion
Anisotropy in a Transverse Magnetic Field}
\author{Huaizhong Xing,$^1$ Gang Su,$^{2,*}$ Song Gao$^3$, and Junhao Chu$^1$}
\address{$^{1}$National Laboratory for Infrared Physics, Shanghai Institute of\\
Technical Physics, Chinese Academy of Sciences, Shanghai 200083, China\\
$^{2}$Department of Physics, The Graduate School of the Chinese Academy of\\
Sciences, P.O. Box 3908, Beijing 100039, China\\
$^{3}$State Key Laboratory of Rare Earth Materials Chemistry and\\
Applications, PKU-HKU Joint Laboratory on Rare Earth Materials and\\
Bioinorganic Chemistry, Peking University, Beijing 100871, China}
\maketitle

\begin{abstract}
The field-induced transition in one-dimensional $S=1$ Heisenberg
antiferromagnet with single-ion anisotropy in the presence of a transverse
magnetic field is obtained on the basis of the Schwinger boson mean-field
theory. The behaviors of the specific heat and susceptibility as functions
of temperature as well as the applied transverse field are explored, which
are found to be different from the results obtained under a longitudinal
field. The anomalies of the specific heat at low temperatures, which might
be an indicative of a field-induced transition from a Luttinger liquid phase
to an ordered phase, are explicitly uncovered under the transverse field. A
schematic phase diagram is proposed. The theoretical results are compared
with experimental observations.
\end{abstract}



\section{INTRODUCTION}

The study on one-dimensional (1D) magnetic systems began in 1930$^{\prime }$%
s.$^{1}$ Although several decades passed, the low-dimensional
antiferromagnets have still received much renewed attention owing to both
experimental and theoretical efforts in the past few years. On the
theoretical aspect, most recent works associated with methods such as
finite-size scaling,$^{3}$ numerical calculations,$^{4,5}$ Monte Carlo
methods,$^{6}$ analyses of an exactly solvable model,$^{7}$ etc., have been
devoted to verifying Haldane$^{\prime }$s conjecture$^{2}$ that the uniform
integer spin chain with Heisenberg antiferromagnetic interactions is massive
whereas the half-integer spin chain is massless, and much progress towards
understanding Haldane$^{\prime }$s scenario has been made so far.

On the other hand, Haldane$^{\prime }$s conjecture has been confirmed
experimentally in a number of 1D antiferromagnets with spin integer. Quite
recently, Honda et al$^{8,9}$ reported an anomaly of heat capacity on a
single crystal sample of the $S=1$ quasi-1D Heisenberg antiferromagnet (HAF) 
$Ni(C_{5}H_{14}N_{2})_{2}N_{3}(PF_{6})$ (briefly NDMAP) in applied magnetic
fields, and identified such an anomaly as an indication of a field-induced
magnetic long-range ordering. They found an anisotropy in the susceptibility
which can be explained as due to the single-ion anisotropy of $Ni^{2+}$. The
model Hamiltonian they adopted is expressed as

\begin{equation}
H=J\sum_{\langle ij\rangle }{\bf S}_{i}{\bf \cdot }{\bf S}%
_{j}+D\sum_{i}(S_{i}^{z})^{2}-\mu _{B}\sum_{i}{\bf S}_{i}\cdot {\bf \tilde{g}%
}\cdot {\bf h\!,}  \label{1}
\end{equation}
where $J$ is the exchange integral (for NDMAP $J/k_{B}\sim 30K$), $D$ is the
single-ion anisotropy constant, $\mu _{B}$ is the Bohr magneton, and ${\bf 
\tilde{g}}$ is the $g$ tensor with elements $g_{\perp }$ and $g_{\Vert }$
corresponding to the $g$ values perpendicular and parallel to the chain c
axis, respectively.

Motivated by this nice experiment, we have recently discussed such an 1D
system in a longitudinal magnetic field.$^{10}$ As this magnetic system is
expected to reveal quite different behaviors in longitudinal and transverse
applied fields, as manifested by experiments$^{8,9}$, it would be
interesting to pay attention to the case in a transverse magnetic field. It
is thus the purpose of this present article to report thermodynamic
behaviors of this system in the presence of a transverse magnetic field by
means of the modified Schwinger boson mean-field theory {\bf (}SBMFT{\bf )}.
As is well known, the SBMFT works well for the 1D HAF with integer spins,
which can include the effects of quantum fluctuations self-consistently in
the large-$N$ saddle point, where $N$ is the quasiparticle degeneracy. The
Schwinger boson approach is actually a large-$N$ formulation, and the large-$%
N$ limit of Hamiltonian is taken with $\kappa =n_{b}/N$ fixed$^{12,13,14}$
with the constraint $b^{+}b=n_{b}$ at each site for Schwinger bosons. For
large values of $\kappa $ the system is magnetically ordered in high
dimensions, while the quantum disordered state appears at small $\kappa $.
In the presence of a magnetic field one may expect the degeneracy to be
lifted and $N$ could be smaller than in the absence of a magnetic field. By
noting that the degeneracy is for Schwinger boson quasiparticles, the SBMFT
still works even if the applied magnetic field lifts some degeneracies,
because even for small $N$ it is known that the SBMFT can produce fair
results for the 1D HAF with integer spin (see, e.g. Ref.[12]). Our
so-obtained results would thus be reliable at the mean-field level.

The outline of this paper is as follows. In next section,{\rm \ }the{\rm \ }%
thermodynamic quantities such as the specific heat and the transverse
susceptibility will be calculated based on the framework of the SBMFT. The
detail formalism of SBMFT for this present system is collected in the
Appendix. Finally, a summary and discussion will be presented.

\section{THERMODYNAMIC PROPERTIES AT FINITE TEMPERATURES}

Based on the self-consistent equations developed in the Appendix, we may
obtain the static uniform transverse susceptibility (along $x$ direction)
per site by 
\begin{eqnarray}
\chi /2N &=&\frac{\mu _{B}^{2}g_{\perp }^{2}}{2N}\sum_{k}\{\beta \frac{%
U^{2}(k)}{1-U^{2}(k)}\{[n_{\alpha }(k)+1]n_{\alpha }(k)+[n_{\beta
}(k)+1]n_{\beta }(k)\}  \nonumber  \label{14b} \\
&&\ +\frac{(1-U^{2}(k))^{-\frac{3}{2}}}{2\Lambda ^{\prime }}(n_{\alpha
}(k)+n_{\beta }(k)+1)\},
\end{eqnarray}
where

\begin{equation}
U(k)=\eta \gamma _{k}-\frac{\mu _{B}g_{\perp }h_{x}}{2\Lambda ^{\prime }},\;%
\text{\ }n_{\alpha (\beta )}(k)=\frac{1}{e^{\beta E_{k}^{\alpha (\beta )}}-1}%
.
\end{equation}
The specific heat per site, $C/2N$, can be easily obtained by $\frac{C(T)}{2N%
}=\frac{1}{2N}\frac{\partial E}{\partial T}$, where $E$ is the internal
energy. For convenience in numerical calculations, we shall set $J=1$ as
energy scale, $g_{\perp }=2.17$ (from Ref. [8]) and $k_{B}=1$ hereafter.

\subsection{Specific Heat}

The temperature dependence of the specific heat ($C_{v}$) for different
fields at $D=0$ is shown as in Fig. 1. One may see that the specific heat
increases with increasing field at low temperature, whereas it decreases
with increasing field at high temperature. At a given magnetic field $%
0<h_{x}<0.9$, there is an anomaly clearly observed in the curve of $C_{v}$
at a temperature $T_{c}$. If the transverse field is taken off, the anomaly
disappears. Such an anomaly might be an indicative of field-induced
transition. Recall that in the presence of a longitudinal field, the system
shows no any anomaly in the curve of $C_{v}$ observed.$^{10}$ In this sense,
this kind of field-induced transition appears only when a transverse
magnetic field is applied to the system within the framework of the SBMFT.
By noting that the anomalies were also observed in experiments for different
longitudinal fields at almost the same temperature, although with smaller
peaks than applying the transverse fields. The fact that our SBMFT cannot
produce this result shows that there might be some limitations for this
approach. With increasing the magnetic field, the temperature $T_{c}$ where
the field-induced transition occurs, moves to low temperature side.
Combining the results of susceptibility (see below), we may identify the
transition as one from a disordered phase to a spin-polarized phase. We
could understand this phenomenon as a consequence of the broken of the $%
Z_{2} $ symmetry.$^{16}$ When the external field is applied along the $z$%
-axis, the $XY$ symmetry is retained. On the contrary, when the magnetic
field is applied perpendicularly to the $z$-axis, the $XY$ symmetry is
broken and an Ising anisotropy is produced (See, e.g. Ref. [8] and
references therein). At a given field an anomaly of specific heat will thus
appear at a given temperature due to the $Z_{2}$ symmetry breaking, which is
an indicative of a field-induced transition from a disordered phase to a
spin-polarized phase as the transverse field tends to align spins
perpendicular to the easy axis. Here we would like to point out that the
anomaly we observed is different from the usual broad peak of $C_{v}$ which
appears at temperature $T_{p}$ whatever the magnetic field is present or
absent, while $T_{p}$ decreases with increasing the field. Since the
limitation of the approximation we cannot find such a broad peak of $C_{v}$
within the frame of SBMFT, as discussed in a number of references. This may
be a shortcoming of the SBMFT. However, at high temperature we have observed
the specific heat decreases with increasing the field, which shows a correct
trend for $S=1$ AFM chain. On the other hand, we note that the anomalies in
the curves of $C_{v}$ were observed experimentally for NDMAP at high
magnetic fields. Our calculated result is in qualitatively agreement with
this experimental obervation$^{8}$ if we make a rescale for the field as $%
h_{x}\longrightarrow (h_{c}-h_{x})$,$^{19}$ where $h_{c}$ is a critical
field at which the anomaly of specific heat disappears at zero temperature.
In the present case $h_{c}\approx 0.90$. With this field rescaling, we find
that the temperature where the anamoly occurs increases with the field,
consistent with the experimental observation. We would like to point out
that we could not compare the experimental data directly with our calculated
results, because the experimental data$^{8}$ include the contribution of the
lattice which is not given. However, one may see that the results obtained
on the basis of the SBMFT in the present fashion might capture qualitatively
some experimental features of the $S=1$ HAF chain in the presence of a
transverse field. Besides, our SBMFT result seems also to be in
qualitatively agreement with the upper phase boundary of the exact
diagonalization of finite chains, although the latter results are extracted
from the staggered susceptibility.$^{16}$

We have also investigated the effect of the single-ion anisotropy on the
specific heat in a given field $h_{x}=0.3$, as shown in Fig. 2. The result
shows that the anisotropy does not have so much effect on the specific heat,
and it only causes slight changes in the peaks and dips of $C_{v}.$ It
appears that $C_{v}$ increases, though small, with increasing $D$, as
indicated in the inset of Fig. 2. One may notice that this result is very
different from the case in the presence of a longitudinal applied field,
where the effect of the single-ion anisotropy on specific heat is various.$%
^{10}$ This result is readily understandable, because the single-ion
anisotropy is XY-like, and if the magnetic field is applied perpendicular to
the $z$-axis, the single-ion anisotropy would have no much effect on the
physical obervables, as illustrated in the present case, while such an
anisotropy would matter if the magnetic field is applied along the $z$-axis,
as manifested in Ref.[10].

The external field dependence of the specific heat for different
temperatures at $D=1.0$ is depicted in Fig. 3. It is shown that at a given
temperature the specific heat increases with increasing field, and after
arriving at a maximum, decreases faster and then slow with increasing field.
One may see again that there are anomalies in the curves of the specific
heat versus the external field. It is interesting to note that the magnetic
field at which the specific heat shows a peak at a given temperature is
consistent with those found in Fig. 1. For a given temperature, $%
C_{v}(h_{x}) $ behaves as $C_{v}(h_{x})\sim
c_{0}(T)+c_{1}(T)h_{x}^{1/2}+c_{2}(T)h_{x}+c_{3}(T)h_{x}^{3/2}$, where $%
c_{i} $ ($i=0,1,2,3$ ) are temperature-dependent coefficients. For example,
at $T=0.5$ the fitting results give $%
c_{0}=0.1643,c_{1}=0.7013,c_{2}=-1.2047,c_{3}=0.468.$ At $%
h_{x}\longrightarrow 0,C(T=0.5)$ is $0.1643$. It can be seen that after
making a field recaling $h_{x}\longrightarrow h_{c}-h_{x}$ one may find that
the position of the anomaly of the specific heat moves to high field side
with increasing temperature.

\subsection{Transverse Susceptibility}

The temperature dependence of susceptibility ($\chi $) at $D=1.0$ in
different fields is depicted in Fig. 4. It is observed that the
susceptibility shows a broad peak at low temperature in low fields (in our
case $h_{x}<0.485$), which is a characteristic of $S=1$ Heisenberg AFM
chain, and it goes to zero at $T\rightarrow 0$ for the field less than $%
h_{c1}=0.485$, suggesting that the system is in a Haldane gapped phase in
this regime. For $0.485<h_{x}<0.9,$ we found the susceptibility of this 1D\
spin system goes to a finite value as $T$ tends to zero, implying that in
this regime the system is in a Luttinger liquid (LL) phase. It is clear that
there exists a transition from the Haldane gapped phase to the LL phase at $%
h_{x}=h_{c1}$ at which the excited gap closes. The result agrees with that
of the effective-field theory and size scaling analysis$^{18}$ with the
finite chain calculation$^{19,20}$, which indicates that the field-induced
transition occurs from the disordered ground state to the gapless
Tomonaga-Luttinger liquid phase at some critical external field $h_{c1}$. In
addition, a phase boundary in field ($h_{c1}$)-temperature ($T$) plane is
thus obtained by observing the variation of susceptibility. On the other
hand, for $h_{c1}<h<h_{c2}$, the gap vanishes and we find a range of linear $%
T$ dependence for the specific heat, which is also a characteristic for the
LL phase.$^{21,22}$ For the higher field ($h_{x}\geq h_{c2}=0.9$) the
susceptibility diverges and persists up to saturation as temperature tends
to zero, showing that the system now enters into a spin-polarized phase.
Thus, there must be a transition from the LL phase to a spin-polarized phase
at $h_{x}=h_{c2}$. However, it is not easy to determine $h_{c2}$ according
to the divergence of the susceptibility. Fortunately, we observed that there
is a curvature change in the curves of $\chi $ versus temperature for
different fields, as enlarged in the inset of Fig. 4. Although the curvature
change in the susceptibility itself is not an indication of a phase
transition, by observing that the susceptibility shows different behaviors
at low temperatures in different phases, one could identify when the system
enters into the spin-polarized phase from the LL phase by observing the
curvature change of the susceptibility. The positions of curvature change
are the same as the positions in curves of the specific heat at which the
specific heat exhibits an anomaly, which gives the estimation of $h_{c2}$.
In other words, at $h_{c2}$ the specific heat shows anomalies, indicating a
transition occurs at $h_{c2}$. Note that the curvature-changing position
moves to low temperature side with increasing field. It can be observed that
the behavior of susceptibility in the transverse field is very different
from the results in the longitudinal field,$^{10}$ where there is no
curvature-changing observed in the curves of $\chi $ versus temperature.

The effect of the single-ion anisotropy on susceptibility is also
investigated. It is found that the single-ion anisotropy does not change so
much the shape and the magnitude of susceptibility. Though the effect on
susceptibility is not obvious, it can cause a small decreasing of $\chi $ at
a given temperature and a given field. One can obtain a schematic phase
diagram in field ($h_x$)-anisotropy ($D$) plane by observing the variation
of susceptibility, as shown in Fig. 5. The phase boundary is characterized
by closure of the excited gap. The system is in a gapless phase above the
boundary and is in a gapped Haldane phase below the boundary. One may note
that the change of $h_x$ with $D$ in the diagram is only a few percent.
Whatever the anisotropy is strong or weak, the system is always in the
Haldane phase if $h_x$ is less than $0.45$. When $D$ is larger, the external
field required to drive the system into the gapless phase becomes larger.
For example, at $D=0,$ the critical field is $0.45$, and at $D=1.6,$ it
becomes $0.50$. It is found that the critical field in the case of the
transverse field is larger than that in the longitudinal field. For
instance, at $D=0$ and $T\rightarrow 0$ the critical field in the transverse
case is $0.45$, while it is $0.155$ in the longitudinal case. Our result is
in agreement with the estimation on the critical field at $D=0$ on the basis
of exact diagonalization of finite chains,$^{16}$ and is qualitatively
consistent with the experimental extrapolation in NDMAP$^8$.

Fig. 6 gives the field-dependence of susceptibility at $D=1.0$ at different
temperatures. It is observed that for $h_x<0.9$ the behavior of $\chi $
versus transverse field at low temperature is quite different from that at
high temperature. The positions of curvature change in $\chi $ are
consistent with our observation for the anomalies in specific heat. We also
observed that the single-ion anisotropy does not have much effect on the
behavior of the susceptibility versus transverse field.

\section{SUMMARY and DISCUSSION}

By summarizing the above results on the specific heat and the
susceptibility, we may propose a schematic phase diagram in field ($h_{x}$%
)-temperature ($T$) plane within the framework of SBMFT. Since the
anisotropy does not have much effect on the behavior of thermodynamic
obervables of the system under interest, without loss of generality we
present the phase diagram for $D=0$ as an example, as shown in Fig. 7. When
the transverse field is less than $h_{c1}$, the system should be in a gapped
phase (Haldane phase). At $h_{x}=h_{c1}$ the gap closes. The lower boundary
for $h_{c1}$ is determined by observing the closure of spin gap in the
curves of the susceptibility as a function of the applied transverse field
and temperature. For $<h_{x}<h_{c2}$, the system might be in a Luttinger
liquid phase, characterized by finite values of the susceptibility at $%
T\longrightarrow 0$. When $h_{x}>h_{c2}$, the system goes into a
spin-polarized phase. At $h_{x}=h_{c2}$, the anomalies in the curves of the
specific heat as well as the susceptibility appear. It is interesting to
observe that the shape of the LL phase becomes more symmetric for $D$
nonzero. It seems that our proposed phase diagram based on the SBMFT is
somewhat different from the results of exact diagonalization on finite
chains.$^{16}$ In the phase diagram presented in Ref. [16] the lower and
upper boundaries of the antiferromagnetic ordered phase are given by $h_{c1}$
and $h_{c2}$, which are determined by the staggered susceptibility. While in
our proposed phase diagram, $h_{c1}$ and $h_{c2}$ are determined by
observing the behaviors of the uniform transverse susceptibility as well as
the specific heat. Considering the mean-field feature of our method, we
cannot determine with accuracy whether a small region of an ordered phase at
very low temperatures ($T<0.1$) in the LL phase exists, as the
susceptibility we calculated is not a staggered but uniform one. In this
sense, our proposed phase diagram is not incompartible with that of
Ref.[16], which can be viewed as a complement for finite-size calculations.
Therefore, it could be better to understand the thermodynamic properties of
this 1D spin system in the presence of the applied field based on a
combination of our SBMFT results and finite-size calculations.

\acknowledgements
This work is supported in part by the National Science Foundation of China
(Grant No. 90103023 and 10104015), the State Key Project for Fundamental
Research in China, and by the Chinese Academy of Sciences.

\appendix

\section{Formalism of the SBMFT}

In this Appendix, we shall present the detail derivation of the
self-consistent equations based on the SBMFT. These equations are applied to
get the thermodynamic properties of the system.

We suppose that the system in Eq.(1) with $J>0$ is defined on a bipartite
lattice with sublattices A and B. For sublattice A, we denote the spin
operator $S_{i}^{A}$ by Schwinger bosons $a_{i}$ and $b_{i}$ (see Ref.
[11]): 
\begin{equation}
\begin{array}{c}
S_{i,+}^{A}=-a_{i}^{+}b_{i}^{+},S_{i,-}^{A}=-a_{i}b_{i}, \\ 
S_{i,z}^{A}=\frac{1}{2}(a_{i}^{+}a_{i}+b_{i}^{+}b_{i}+1),
\end{array}
\stackrel{}{\qquad i\in A}  \label{2a}
\end{equation}
satisfying the constraint $S_{i}^{A}=(a_{i}^{+}a_{i}+b_{i}^{+}b_{i})$ for
each site on sublattice A, and for sublattice B

\begin{equation}
\begin{array}{r}
S_{j,+}^{B}=a_{j}b_{j},S_{j,-}^{B}=a_{j}^{+}b_{j}^{+} \\ 
S_{j,z}^{B}=-\frac{1}{2}(b_{j}^{+}b_{j}+a_{j}^{+}a_{j}+1)
\end{array}
\stackrel{}{\qquad j\in B}  \label{2b}
\end{equation}
with $S_{j}^{B}=(a_{j}^{+}a_{j}+b_{j}^{+}b_{j})$ for each site on sublattice
B. The bosons $\{a\}$ and \{$b\}$ obey the standard commutation relations.
Although one may note that the definitions in Eqs. (2) and (3) introduced in
Ref. [11] are slightly different from the standard form, we find that such a
form is quite convenient for our purpose. On account of the definitions with
constraints, the Hamiltonian (1) for $S=1$ can be rewritten as

\begin{equation}  \label{3}
H=H_0+H_D+H_{ex},
\end{equation}

\begin{equation}
\begin{array}{c}
H_{0}=J\sum_{\langle i,j\rangle }{\bf S}_{i}\cdot {\bf S}_{j}=-2J\sum_{i=1,%
\delta }A_{i,i+\delta }^{+}A_{i,i+\delta } \\ 
-J\sum_{i=1,\delta }a_{i}^{+}a_{i}a_{i+\delta }^{+}a_{i+\delta
}+J\sum_{i=1,\delta }a_{i}^{+}a_{i},
\end{array}
\end{equation}

\begin{equation}
H_D=D\sum_{i=1}(S_i^z)^2=\frac{3D}4\sum_{i=1}(a_i^{+}a_i+b_i^{+}b_i)+DN/2,
\label{3b}
\end{equation}

\begin{equation}
H_{ex}=-\mu _B\sum_i[g_{\perp }(S_i^xh_x+S_i^yh_y)+g_{\Vert }S_i^zh_z],
\label{3c}
\end{equation}
where $A_{i,i+\delta }=\frac 12(a_i^{+}b_{i+\delta }^{+}+b_ia_{i+\delta })$,
and $2N$ is the total number of sites. To implement the constraints $%
\sum_{i\in A}(a_i^{+}a_i+b_i^{+}b_i)=S^A$ and $\sum_{j\in
B}(a_j^{+}a_j+b_j^{+}b_j)=S^B$ where $S^{A(B)}$ denotes the total spin of
sublattice A(B), we should introduce two kinds of Lagrangian multipliers $%
\lambda _i^A$ and $\lambda _j^B$ into the system. At the mean-field level,
for the sake of simplicity we may take the average value of the bond
operator $<A_{i,i+\delta }>=A$ to be uniform and static, and so are $%
<\lambda _i^A>=$ $\lambda ^A$ and $<\lambda _j^B>=\lambda ^B$.

The mean-field Hamiltonian in momentum space reads 
\begin{equation}
H=H_0+H_D+H_{ex},  \label{5}
\end{equation}
\begin{equation}
\begin{array}{c}
H_0=-J\sum_k\{A^{*}(a_{-k}^{+}b_k^{+}+b_ka_{-k})z\gamma
_k^{*}+(b_ka_{-k}+a_{-k}^{+}b_k^{+})z\gamma _kA\} \\ 
-J\sum_k(<\widetilde{n}_{i+\delta ,a}>a_{-k}^{+}a_{-k}+<\widetilde{n}%
_{i,a}>a_{-k}^{+}a_{-k}) \\ 
+J\sum_ka_{-k}^{+}a_{-k}+\lambda ^A\sum_k[(a_{-k}^{+}a_{-k}+b_k^{+}b_k)-S^A]
\\ 
+\lambda ^B\sum_k[(a_{-k}^{+}a_{-k}+b_k^{+}b_k)-S^B] \\ 
+2JzNA^{*}A+JNz<\widetilde{n}_{i,a}><\widetilde{n}_{i+1,a}>,
\end{array}
\label{5a}
\end{equation}
\begin{equation}
H_D=\frac{3D}4\sum_{i=1}(a_{-k}^{+}a_{-k}+b_k^{+}b_k)+DN/2,  \label{5b}
\end{equation}
\begin{equation}
H_{ex}=-\frac 12\mu _B\sum_k[-g_{\perp }h_x(a_{-k}^{+}b_k^{+}+b_ka_{-k})+%
\frac{-1}jg_{\perp }h_y(a_{-k}^{+}b_k^{+}-a_{-k}b_k)+g_{\Vert
}h_z(a_{-k}^{+}a_{-k}+b_k^{+}b_k+1)],  \label{5c}
\end{equation}
where $z$ is the number of nearest neighbor sites, $j$ is a complex number,
and $\gamma _k=\frac 1z\sum_\delta e^{ik\delta }=\cos k$. The sum over $k$
is restricted to the reduced first Brillouin zone.

Utilizing the Bogoliubov transformation 
\begin{equation}
a_{-k}=\cosh \theta _{k}\alpha _{k}+\sinh \theta _{k}\beta
_{k}^{+},\;b_{k}=\sinh \theta _{k}\alpha _{k}^{+}+\cosh \theta _{k}\beta
_{k},\quad  \label{6}
\end{equation}
with $\theta $ given by \qquad 
\begin{equation}
tanh2\theta =\frac{-\mu _{B}g_{\perp }h_{x}+Re(2JzA\gamma _{k})}{2\Lambda
^{^{\prime }}},  \label{7a}
\end{equation}
with 
\[
2\Lambda ^{^{\prime }}=(\lambda ^{A}+\lambda ^{B})-B/2+3D/4,\qquad B\equiv
J[(<\widetilde{n}_{i+1,a}>+<\widetilde{n}_{i,a}>)-1)], 
\]
we obtain the energy spectrum of the system in the presence of a transverse
external field $h_{ex}=(h_{x},0,0)$: 
\begin{equation}
E_{k}^{\alpha }=\sqrt{(2\Lambda ^{^{\prime }})^{2}-(2Re\left| JzA\gamma
_{k}\right| -\mu _{B}g_{\perp }h_{x})^{2}}-B/2,  \label{7b}
\end{equation}
\begin{equation}
E_{k}^{\beta }=\sqrt{(2\Lambda ^{^{\prime }})^{2}-(2Re\left| JzA\gamma
_{k}\right| -\mu _{B}g_{\perp }h_{x})^{2}}+B/2.  \label{7c}
\end{equation}
The free energy per site takes the form \negthinspace of 
\begin{equation}
\begin{array}{c}
\frac{F}{2N}=\frac{1}{2N\beta }\sum_{k}\{\ln [2\sinh \frac{\beta }{2}%
(E_{k}^{\alpha })]+\ln [2\sinh \frac{\beta }{2}(E_{k}^{\beta })]\} \\ 
+JzA^{\ast }A+B/4-(S^{A}/2+1/2)\lambda ^{A}-\lambda ^{B}(S^{B}/2+1/2) \\ 
-D/2+Jz<\widetilde{n}_{i,a}><\widetilde{n}_{i+1,a}>/2.
\end{array}
\label{8b}
\end{equation}
The mean-field self-consistent equations can be obtained by minimizing the
free energy. Without loss of generality, we may set $\lambda ^{A}=\lambda
^{B}$. Then, we perform $\delta F/\delta \lambda ^{A}=0$, $\delta F/\delta
A^{\ast }=0$, $\delta F/\delta <\widetilde{n}_{i+\delta ,a}>=0$ and $\delta
F/\delta <\widetilde{n}_{i,a}>=0$. Rescale the parameters $(\lambda ^{A},$ $%
\lambda ^{B},A,$ $\beta )$ $\longrightarrow (\Lambda ,\eta ,\kappa ):$ 
\begin{equation}
\eta =\frac{JAz}{\Lambda ^{^{\prime }}},\text{ \quad }\beta =\frac{4\kappa }{%
z}.
\end{equation}
Then, the angle in the Bogoliubov transformation can be expressed in a
compact form \qquad 
\begin{equation}
\cosh 2\theta _{k}=\frac{1}{\sqrt{1-(\eta \gamma _{k}-\frac{\mu _{B}g_{\perp
}h_{x}}{2\Lambda ^{\prime }})^{2}}},\qquad \sinh 2\theta _{k}=\frac{-\frac{%
\mu _{B}g_{\perp }h_{x}}{2\Lambda ^{\prime }}+\eta \gamma _{k}}{\sqrt{%
1-(\eta \gamma _{k}-\frac{\mu _{B}g_{\perp }h_{x}}{2\Lambda ^{\prime }})^{2}}%
}.  \label{12}
\end{equation}
The self-consistent equations become 
\begin{equation}
\begin{array}{c}
(S^{A}+S^{B})/2+1=\frac{1}{2N}\sum_{k}\frac{1}{\sqrt{1-(\eta \gamma _{k}-%
\frac{\mu _{B}g_{\perp }h_{x}}{2\Lambda ^{\prime }})^{2}}}\{\coth \kappa
(2\Lambda ^{\prime }\sqrt{1-(\eta \gamma _{k}-\frac{\mu _{B}g_{\perp }h_{x}}{%
2\Lambda ^{\prime }})^{2}}+B/2) \\ 
+\coth \kappa (2\Lambda ^{\prime }\sqrt{1-(\eta \gamma _{k}-\frac{\mu
_{B}g_{\perp }h_{x}}{2\Lambda ^{\prime }})^{2}}-B/2)\},
\end{array}
\label{13a}
\end{equation}

\begin{equation}
\begin{array}{c}
\Lambda ^{\prime }=\frac{1}{2N}\sum_{k}\frac{J\gamma _{k}}{\eta }\cdot \frac{%
-\frac{\mu _{B}g_{\perp }h_{x}}{2\Lambda ^{\prime }}+\eta \gamma _{k}}{\sqrt{%
1-(\eta \gamma _{k}-\frac{\mu _{B}g_{\perp }h_{x}}{2\Lambda ^{\prime }})^{2}}%
}\{\coth \kappa (2\Lambda ^{\prime }\sqrt{1-(\eta \gamma _{k}-\frac{\mu
_{B}g_{\perp }h_{x}}{2\Lambda ^{\prime }})^{2}}+B/2) \\ 
+\coth \kappa (2\Lambda ^{\prime }\sqrt{1-(\eta \gamma _{k}-\frac{\mu
_{B}g_{\perp }h_{x}}{2\Lambda ^{\prime }})^{2}}-B/2)\},
\end{array}
\label{13b}
\end{equation}
\begin{equation}
\begin{array}{c}
4<\widetilde{n}_{i,a}>+1= \\ 
\frac{1}{2N}\sum_{k}\{[\frac{1}{\sqrt{1-(\eta \gamma _{k}-\frac{\mu
_{B}g_{\perp }h_{x}}{2\Lambda ^{\prime }})^{2}}}-1]\coth \kappa (2\Lambda
^{\prime }\sqrt{1-(\eta \gamma _{k}-\frac{\mu _{B}g_{\perp }h_{x}}{2\Lambda
^{\prime }})^{2}}+B/2) \\ 
+[\frac{1}{\sqrt{1-(\eta \gamma _{k}-\frac{\mu _{B}g_{\perp }h_{x}}{2\Lambda
^{\prime }})^{2}}}-1]\coth \kappa (2\Lambda ^{\prime }\sqrt{1-(\eta \gamma
_{k}-\frac{\mu _{B}g_{\perp }h_{x}}{2\Lambda ^{\prime }})^{2}}-B/2)\},
\end{array}
\label{13c}
\end{equation}

\begin{equation}
\begin{array}{c}
4<\widetilde{n}_{i+1,a}>+1= \\ 
\frac{1}{2N}\sum_{k}\{[\frac{1}{\sqrt{1-(\eta \gamma _{k}-\frac{\mu
_{B}g_{\perp }h_{x}}{2\Lambda ^{\prime }})^{2}}}+1]\coth \kappa (2\Lambda
^{\prime }\sqrt{1-(\eta \gamma _{k}-\frac{\mu _{B}g_{\perp }h_{x}}{2\Lambda
^{\prime }})^{2}}-B/2) \\ 
+[\frac{1}{\sqrt{1-(\eta \gamma _{k}-\frac{\mu _{B}g_{\perp }h_{x}}{2\Lambda
^{\prime }})^{2}}}-1]\coth \kappa (2\Lambda ^{\prime }\sqrt{1-(\eta \gamma
_{k}-\frac{\mu _{B}g_{\perp }h_{x}}{\Lambda })^{2}}+B/2)\}.
\end{array}
\label{13d}
\end{equation}
As is easily seen, the aforementioned equations are highly coupled, and it
is not possible to get useful analytic results. They are only solved
numerically to get the relavant thermodynamic quantities.

\newpage

\begin{center}
FIGURE\ CAPTIONS
\end{center}

Fig. 1 The temperature dependence of the specific heat at $D=0$ for
different transverse fields.

Fig. 2 The temperature dependence of the specific heat for different
anisotropies at $h_{x}=0.3$. Inset: The enlarged part of the specific heat
versus temperature for different anisotropies at $h_{x}=0.3$.

Fig. 3 The $h_{x}$-dependence of the specific heat at $D=1.0$ for different
temperatures.

Fig. 4 The temperature dependence of the susceptibility at $D=1.0$ for
different transverse fields. Inset: The enlarged part of the susceptibility
versus temperature at $D=1.0$ for different transverse fields.

Fig. 5 Schematic phase diagram in $h_{x}-D$ plane. The solid line is the
phase boundary on which the gap closes.

Fig. 6 The $h_{x}$-dependence of the susceptibility at $D=1.0$ for different
temperatures.

Fig. 7 Schematic phase diagram in $h_{x}-T$ plane.

\end{document}